\documentclass[aps,pra,twocolumn,showpacs,amsmath,amssymb,a4paper]{revtex4}

\usepackage{amsthm}
\usepackage{amsmath}
\usepackage{latexsym}
\usepackage{amsfonts}
\usepackage{amssymb}
\usepackage{color}
\usepackage{bbm,dsfont}
\usepackage{graphicx}
\usepackage{mathrsfs}
\usepackage{mathbbol}

\newtheorem{theorem}{Theorem}

\theoremstyle{definition}

\newcommand{\real}{\mathbb R} 
\newcommand{\integer}{\mathbb Z} 
\newcommand{\NAT}{\mathbb N} 
\newcommand{\half}{\tfrac{1}{2}} 
\newcommand{\mo}[1]{\left| #1 \right|} 

\newcommand{\hi}{\mathcal{H}} 
\newcommand{\lh}{\mathcal{L(H)}} 
\newcommand{\trh}{\mathcal{T(H)}} 
\newcommand{\ip}[2]{\left\langle\,#1\,|\,#2\,\right\rangle} 
\newcommand{\kb}[2]{|#1\rangle\langle#2|} 
\newcommand{\tr}[1]{\textrm{tr}\left[#1\right]} 
\newcommand{\id}{\mathbbm{1}} 

\newcommand{\Ao}{\mathsf{A}}
\newcommand{\Bo}{\mathsf{B}}
\newcommand{\Mo}{\mathsf{M}}
\newcommand{\No}{\mathsf{N}}
\newcommand{\To}{\mathsf{T}}
\newcommand{\Qo}{\mathsf{Q}}
\newcommand{\Po}{\mathsf{P}}

\newcommand{\Phase}{\Phi}

\newcommand{\jmd}{\mathsf{j}} 

\newcommand{\R}{\mathbb{R}}
\newcommand{\C}{\mathbb{C}}
\newcommand{\hil}{\mathcal{H}}

\newcommand{\M}{\mathsf{M}}

\begin{document}

\title{Maximally Incompatible Quantum Observables}

\author{Teiko Heinosaari$^{1}$, Jussi Schultz$^{2}$, Alessandro Toigo$^{2,3}$, and Mario Ziman$^{4,5}$}


\affiliation{
$^{1}$Turku Centre for Quantum Physics, Department of Physics and Astronomy, University of Turku, FI-20014 Turku, Finland\\
$^{2}$Dipartimento~di~Matematica,~Politecnico~di~Milano,~Piazza~Leonardo~da~Vinci~32,~I-20133~Milano, Italy\\
$^{3}$Istituto Nazionale di Fisica Nucleare, Sezione di Milano, Via Celoria 16, I-20133 Milano, Italy\\
$^{4}$RCQI, Institute of Physics, Slovak Academy of Sciences, D\'ubravsk\'a cesta 9, 84511 Bratislava, Slovakia\\
$^{5}$Faculty of Informatics, Masaryk University, Botanick\'a 68a, 60200 Brno, Czech Republic
}


\begin{abstract}
The existence of maximally incompatible quantum observables in the sense of a minimal joint measurability region is investigated. Employing the universal quantum cloning device it is argued that only infinite dimensional quantum systems can accommodate maximal incompatibility.
It is then shown that two of the most common pairs of complementary observables (position and momentum; number and phase) are maximally incompatible. 
\end{abstract}

\maketitle

\section{Introduction}

One of the peculiar features that one encounters when entering the realm of quantum physics is the impossibility of measuring certain observables jointly with a single measurement setup. This \emph{incompatibility} of observables has various manifestations, captured for instance in the concept of complementarity \cite{BuLa95} or the  uncertainty principle \cite{BuHeLa07}. However, the formulation of these features is not restricted to quantum theory and can be carried out also in a more general framework \cite{BuLa80}. This opens up the possibility of exploring which, if any, of these features are characteristic of quantum theory.  

In a recent Letter \cite{BuHeScSt13} a new way of comparing the incompatibility of pairs of observables using the concept of \emph{joint measurability region} was introduced. The joint measurability region of a pair of observables describes the amount of noise that needs to be added in order to make the observables jointly measurable. This concept is well-defined in any probabilistic theory, thus allowing the comparison even between pairs of observables in different theories. In particular, we can define a \emph{maximally incompatible} pair of observables to be a pair whose joint measurability region is as small as it can be in any probabilistic theory. 

It was demonstrated in \cite{BuHeScSt13} that quantum theory \emph{does} contain maximally incompatible observables, although the provided example was based on a rather artificial construction. The purpose of this Letter is to complement the earlier work \cite{BuHeScSt13} by shedding more light onto the maximally incompatible observables in quantum theory. Firstly, we show that the infinite dimensionality of the Hilbert space is a necessary condition for maximal incompatibility. Secondly, we present physically relevant examples of maximal incompatibility by proving that the canonically conjugated position and momentum observables, as well as the number and phase observables, constitute maximally incompatible pairs. 

\section{Joint measurability degree}

An observable $\Mo$ in quantum theory is generally described by a normalized positive operator valued measure (POVM) \cite{MLQT12}. For the purpose of our investigation, it is sufficient to consider observables whose outcome space $\Omega$ is either $\R^n$ or some subset of $\R^n$.

Two observables $\Mo_1$ and $\Mo_2$ with outcome spaces $\Omega_1$ and $\Omega_2$, respectively, are \emph{jointly measurable} if there exists a third observable $\Mo$ with the product outcome space $\Omega_1\times\Omega_2$ such that $\Mo_1$ and $\Mo_2$ are the margins of $\Mo$, i.e., 
\begin{align*}
\Mo(X\times\Omega_2) = \Mo_1(X) \, , \quad \Mo(\Omega_1\times Y) = \Mo_2(Y) 
\end{align*}
for all Borel sets $X \subseteq \Omega_1$ and $Y\subseteq \Omega_2$.

We say that an observable $\To$ is trivial if $\To(X) = \mu(X) \id$ for some probability measure $\mu$. Hence the obtained measurement outcome does not depend on the input state at all. The fact that a trivial observable is jointly measurable with any other observable serves as a motivation for the following definition \cite{BuHeScSt13}:  For any two observables $\Mo_1$ and $\Mo_2$, the joint measurability region $J(\Mo_1,\Mo_2)$ is the set of all points $(\lambda,\mu)\in[0,1]\times[0,1]$ for which there exist trivial observables $\To_1$ and $\To_2$ such that $\lambda \Mo_1 + (1-\lambda) \To_1$ and $\mu \Mo_2 + (1-\mu) \To_2$ are jointly measurable. It was shown in \cite{BuHeScSt13} that 
the triangle shaped set
\begin{equation*}
\triangle \equiv \{(\lambda,\mu) \in [0,1]\times[0,1]
\vert \lambda+\mu\leq 1\} 
\end{equation*}
is always contained in $J(\Mo_1,\Mo_2)$. This inclusion simply means that once the added noise exceeds a certain bound, then all pairs of observables become jointly measurable. 
Hence, it is natural to say  that two observables $\Mo_1$ and $\Mo_2$ are \emph{maximally incompatible} if their joint measurability region is precisely this minimal set $\triangle$, i.e., $J(\Mo_1,\Mo_2)=\triangle$.

\begin{figure}
\begin{center}
\includegraphics[width=4cm]{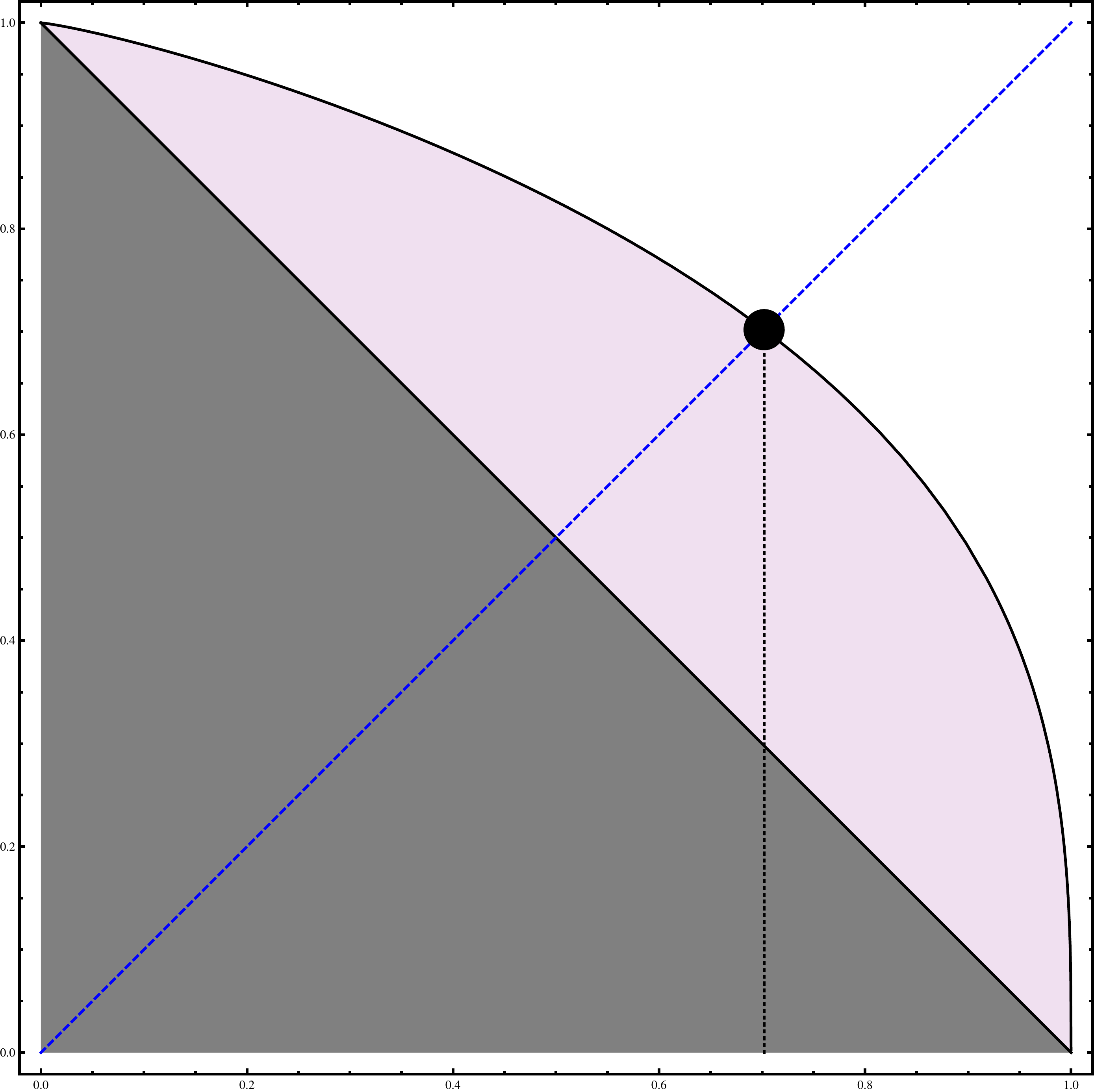} 
\end{center}
\caption{\label{fig:region}(Color online) The region $\triangle$ (dark) is always a subset of the joint measurability region (colored) of two observables, and is equal to it for maximally incompatible observables. The joint measurability degree is graphically obtained as the $\lambda$-coordinate of the intersection (dot) of the boundary of the joint measurability region and the symmetry line $\lambda=\mu$ (dashed blue line).}
\end{figure}

For two observables $\Mo_1$ and $\Mo_2$, we denote by $\jmd(\Mo_1,\Mo_2)$ the greatest number $0\leq \lambda\leq 1$ such that $(\lambda,\lambda)\in J(\Mo_1,\Mo_2)$, and we call it the \emph{joint measurability degree} of $\Mo_1$ and $\Mo_2$ (see Fig.~\ref{fig:region}). 
The joint measurability degree can be seen as another expression of the incompatibility of two observables, coarser than the joint measurability region \cite{StBu13}. (Related concepts have been used also in \cite{BaGaGhKa13, Gudder13}.)
Note that $\half\leq\jmd(\Mo_1,\Mo_2) \leq 1$ since $\triangle\subseteq J(\Mo_1,\Mo_2) \subseteq [0,1]\times[0,1]$, and the convexity of $J(\Mo_1,\Mo_2)$ implies that \emph{$\Mo_1$ and $\Mo_2$ are maximally incompatible if and only if $\jmd(\Mo_1,\Mo_2)=\half$}.  

In a finite $d$-dimensional Hilbert space $\hi_d$ a natural candidate for a maximally incompatible pair is the canonically conjugated pair corresponding to two mutually unbiased bases that are connected via finite Fourier transform \cite{Kraus87}. Fix an orthonormal basis $\{\varphi_j\}_{j=0}^{d-1}$ of $\hi_d$ and define
\begin{equation}\label{eq:defF}
\psi_k= \frac{1}{\sqrt{d}} \sum_{j=0}^{d-1} e^{2\pi i jk/d} \varphi_j \, .
\end{equation}
It is immediate to check that $\{\varphi_j\}_{j=0}^{d-1}$ and $\{\psi_k\}_{k=0}^{d-1}$ are mutually unbiased, i.e, $\mo{\ip{\varphi_j}{\psi_k}}=1/\sqrt{d}$ for all $j,k$.
The corresponding observables $\Ao(j)=\kb{\varphi_j}{\varphi_j}$ and $\Bo(k)=\kb{\psi_k}{\psi_k}$ are thus complementary in the sense that if $\tr{\varrho\Ao(j)}=1$ for some state $\varrho$, then $\tr{\varrho\Bo(k)}=1/d$, and vice versa.
However, it has been proved in \cite{CaHeTo12} that 
\begin{equation}\label{eq:jmd-mub}
\jmd(\Ao,\Bo)=\frac{2+\sqrt{d}}{2(1+\sqrt{d})} \, , 
\end{equation}
so that $\Ao$ and $\Bo$ are not maximally incompatible. Nevertheless, this does not
rule out the existence of a maximally incompatible pair of observables for finite dimensional systems.
\section{Bounds for the joint measurability degree of finite dimensional observables}
\begin{figure}
\includegraphics[width=4cm]{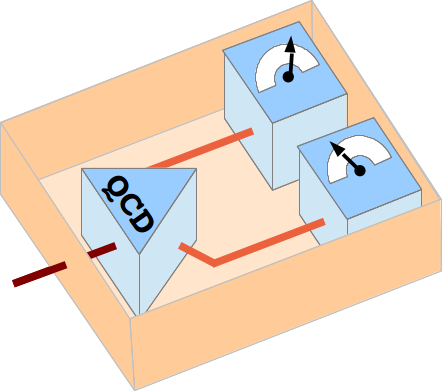}
\caption{(Color online) A quantum cloning device ({\tt QCD}) can be
employed for performing an approximate joint measurement of an arbitrary pair 
of observables by making two approximate clones of the initial state and then performing measurements separately on these clones.}
\label{fig:cloning}
\end{figure}

If perfect cloning of quantum states would be possible, then obviously all observables would be jointly measurable. 
Even if this is not the case, we may try to use an imperfect but realizable cloning device as a way of performing approximate joint measurements.
The method is very simple; we make two approximate clones $\tilde{\varrho}$ of the initial state $\varrho$.
Then we perform measurements of $\Mo_1$ and $\Mo_2$ separately on these two approximate clones; see Fig.~\ref{fig:cloning}.
The resulting total measurement is not a joint measurement of $\Mo_1$ and $\Mo_2$, but of their noisy versions. 
The additional noise clearly depends on the performance of the quantum cloning device.

We consider the cloning device $C$ of the form \cite{KeWe99}
\begin{equation*}
C(\varrho) = \frac{2}{d+1} \, S(\varrho \otimes \id) S\, , 
\end{equation*}
where $S$ is the projection from $\hi_d\otimes\hi_d$ to the symmetric subspace of $\hi_d\otimes\hi_d$.
The state $\tilde{\varrho}$ of each approximate clone is obtained as the corresponding partial trace of $C(\varrho)$ and we get 
\begin{equation*}
\tilde{\varrho} = c(d) \varrho + (1-c(d)) \frac{\id}{d} \, ,
\end{equation*}
where the number $c(d)$ depends only on the dimension $d$ and is given by
\begin{equation*}
c(d) = \frac{2+d}{2(1+d)} \, .
\end{equation*}

For any two observables $\Mo_1$ and $\Mo_2$ we can now define an observable $\Mo$ by the formula
\begin{equation*}
\tr{\varrho \Mo(X\times Y)} = \tr{C(\varrho) \Mo_1(X) \otimes \Mo_2(Y)} \, , 
\end{equation*}
required to hold for all states $\varrho$ and all outcome sets  $X$ and $Y$.
By evaluating  $\tr{\varrho \Mo(X\times \Omega_2)} = \tr{\tilde{\varrho} \Mo_1(X)}$ we obtain the first margin of the observable $\Mo$ as 
$$ 
\Mo(X\times \Omega_2)= c(d)  \Mo_1(X)  + (1-c(d)) \To_1(X)
$$
where the trivial observable $\To_1$ is given by $\To_1(X)=\tr{\Mo_1(X)/d} \id$.
Similarly, 
\begin{align*}
\Mo(\Omega_1\times Y)  = c(d) \Mo_2(Y) + (1-c(d)) \To_2(Y)
\end{align*}
with $\To_2(Y)=\tr{\Mo_2(Y)/d} \id$. We have thus proved the following result.

\begin{theorem}\label{thm:cloning}
Let $\hi_d$ be a $d$-dimensional Hilbert space with $2\leq d<\infty$.
For any two observables $\Mo_1$ and $\Mo_2$ on $\hi_d$ we have
\begin{equation}\label{eqn:cloning_bound}
\frac{1}{2} < \frac{2+d}{2(1+d)} \leq \jmd(\Mo_1,\Mo_2) \, .
\end{equation}
In particular, there are no maximally incompatible observables in a finite dimensional Hilbert space.
\end{theorem}

It is interesting to note that this kind of a restriction  is not a common feature of general probabilistic theories. It was shown in \cite{BuHeScSt13} that there exist theories for which a maximally incompatible pair of observables exists even for the simplest finite system.

\begin{figure}
\begin{center}
\includegraphics[width=5cm]{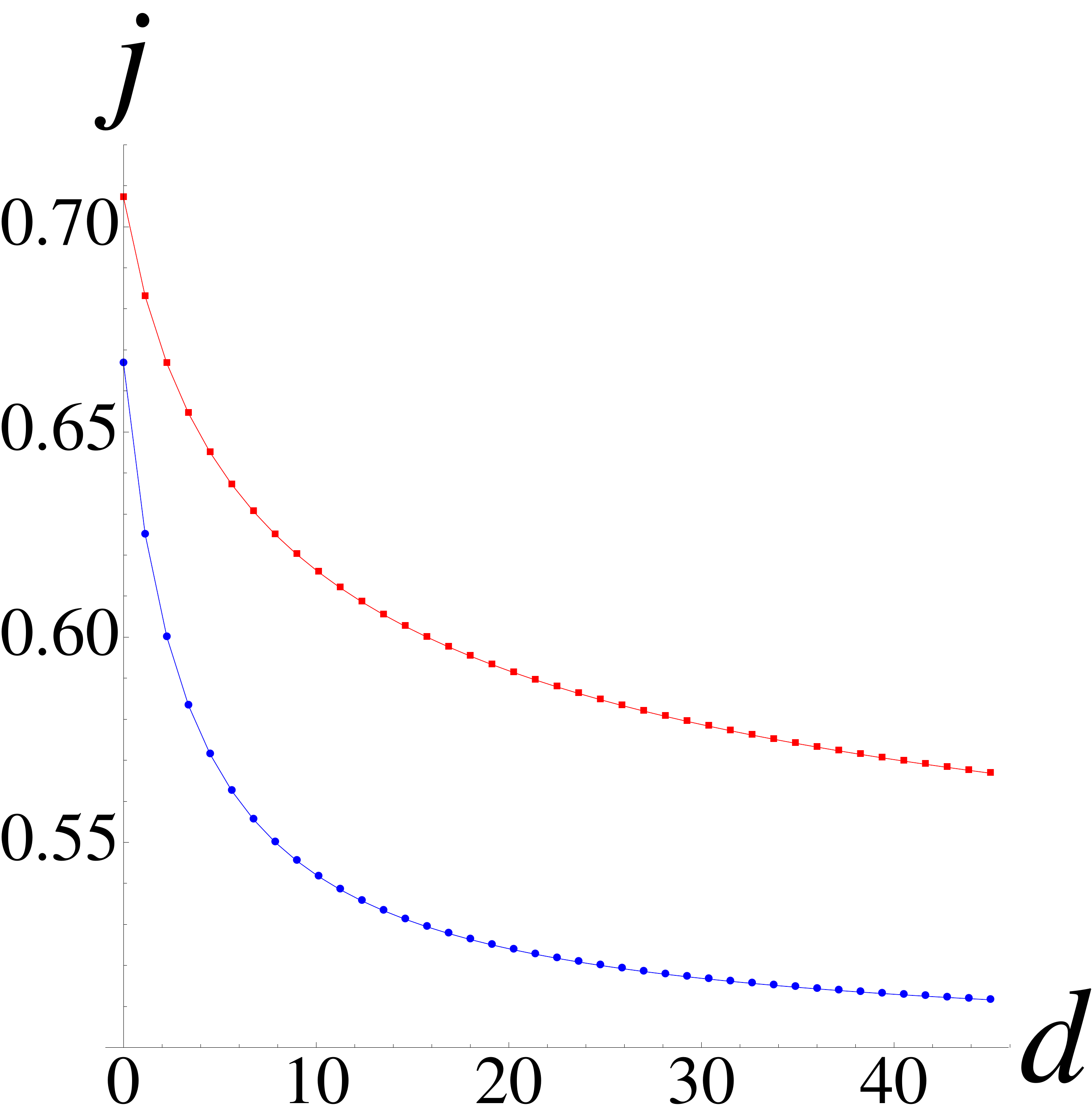}
\end{center}
\caption{(Color online) The joint measurability degree of two canonically conjugated observables (upper curve) and the lower bound obtained by employing a quantum cloning device (lower curve) as functions of the Hilbert space dimension $d$. 
Both of these sequences approach to $\half$ when $d$ goes to infinity, but they are always separated.
The smallest joint measurability degree for two observables in dimension $d$ is somewhere between these curves.
}
\label{fig:comparison}
\end{figure}

Even though Theorem \ref{thm:cloning} gives us a lower bound for the joint measurability degree, it does not tell us whether or not it can actually be reached by any pair of observables. Comparison of Eq.~\eqref{eq:jmd-mub} and Eq.~\eqref{eqn:cloning_bound} immediately implies (see Fig.~\ref{fig:comparison}) that this is not the case for the canonically conjugated pair of observables. That is, the  smallest possible joint measurability degree in a fixed dimension remains an open question.

\section{Maximal incompatibility of position and momentum}
Our first example of a pair of maximally incompatible quantum observables is given by the position and momentum of a particle moving in a single spacial dimension. Consider the Hilbert space $\hil = L^2(\real)$ and the canonical position and momentum observables  $\Qo$ and $\Po$:
\begin{align*}
\ip{\psi}{\Qo(X)\psi}  = \int_X \mo{\psi(x)}^2 dx, \\
\ip{\psi}{\Po(Y)\psi}  = \int_Y \mo{\widehat{\psi}(y)}^2 dy \, , 
\end{align*}
where $\widehat{\psi}$ is the Fourier transform of $\psi$. Similar to the finite dimensional canonical pair $\Ao$ and $\Bo$, also position and momentum are (probabilistically) complementary in the sense that for any bounded intervals $X,Y \subset\real$ and any state $\varrho$, $\tr{\varrho \Qo(X)}=1$ implies $0<\tr{\varrho \Po(Y)}<1 $ and vice versa. It follows that any positive operator $A$ satisfying $A \leq \Qo(X)$ and $A\leq\Po(Y)$ is necessarily zero \cite[Sec.~IV.2.3, IV.2.4]{OQP97}. Using complementarity and the specific structure of $\Qo$ and $\Po$, we can prove the following result.

\begin{theorem}\label{thm:qp}
The position and momentum observables are maximally incompatible.
\end{theorem}

Before going into the details of the proof of this result, we will explain the used main tool.
The starting point is the fact that position and momentum share specific symmetry properties with respect to phase space translations as represented by the Weyl operators $W(q,p)=e^{i\frac{qp}{2}}e^{-iqP}e^{ipQ}$ where $Q$ and $P$ are the selfadjoint position and momentum operators:
\begin{align}
W(q,p)\Qo(X) W(q,p)^*  &= \Qo(X+q),\label{eqn:pos_symmetry}\\
W(q,p)\Po(Y) W(q,p)^*  &= \Po(Y+p).\label{eqn:mom_symmetry}
\end{align} 
If one wishes to add noise to $\Qo$ and $\Po$ while keeping these symmetry properties, then instead of mixing with trivial observables one should convolve them with probability measures \cite{CaHeTo04}.
These smeared position and momentum observables are jointly measurable if and only if they have a joint observable $\Mo$ which is covariant with respect to phase space translations \cite{CaHeTo05}, i.e., $W(q,p)\Mo(Z) W(q,p)^*= \Mo(Z+(q,p))$. 
The proof of this result is based on averaging the joint observable with respect to phase space translations. 
However, since $\R^2$ is not compact  one needs to be careful how to perform this averaging. 
Indeed, one should do this using an invariant mean \cite{Werner04}, which is also the main tool in our proof of Theorem \ref{thm:qp}.

An invariant mean on $B(\R^n)$, the space of bounded complex valued functions on $\R^n$, is a positive linear functional $m:B(\R^n)\to\C$ which is normalized to $m(1) = 1$ and which is invariant with respect to translations (for the existence of invariant means, see \cite[Thm.~17.5]{AHAI63}). More explicitly, if $f^x$ denotes the translate of $f$, i.e., $f^x(y) = f(y+x)$, then $m(f^x)=m(f)$. 

Any observable $\Mo$ on $\R^2$ can be averaged by the following procedure: 
 For any state $\varrho$ and any $f\in C_b(\R^2)$, the space of bounded continuous functions on $\R^2$, we define the bounded function $\Theta [f;\varrho]$ by
\begin{equation*}
\Theta [f;\varrho](q,p) = \tr{\varrho W(q,p) \Mo[f^{(q,p)}]W(q,p)^*}
\end{equation*}
where $\Mo[f]=\int f\, {\rm d}\Mo$.
Now let $m$ be an invariant mean on $B(\R^2)$. Then by the duality $\trh^*\simeq\lh$ between the trace class and the bounded operators, the formula 
\begin{equation}\label{eq:defMav}
\tr{\varrho \Mo^{\rm av}(f)} = m(\Theta [f;\varrho])
\end{equation}
defines a positive linear map $\Mo^{\rm av}:C_b(\R^2)\to\lh$ which is normalized to $\Mo^{\rm av}(1)=\id$. By the analogue of the Riesz-Markov theorem for operator measures \cite[Thm.~19]{NST66}, the restriction of this map to the subspace $C_c(\R^2)\subset C_b(\R^2)$ of compactly supported functions corresponds to a unique POVM $\Mo_0$ on $\R^2$ via $\Mo^{\rm av}(f) = \Mo_0[f]$ for all $f\in C_c(\R^2)$. Note that $\Mo_0$ is phase space translation covariant 
 since $\Mo^{\rm av} (f^{(q,p)}) = W(q,p)^* \Mo^{\rm av} (f) W(q,p)$. However, the equality $\Mo^{\rm av}(f) = \Mo_0[f]$ need not hold for all $f\in C_b(\R^2)$, although in general one has $\Mo_0 [f] \leq \Mo^{\rm av}(f)$ for all positive $f\in C_b(\R^2)$. Indeed, for such functions,
\begin{align*}
\Mo_0 [f] & = \sup\{ \Mo_0 [gf]  \vert g\in C_c(\R^2),\, 0\leq g\leq 1\} \\
& = \sup\{ \Mo^{\rm av} (gf)  \vert g\in C_c(\R^2),\, 0\leq g\leq 1\} \\
& \leq \Mo^{\rm av} (f).
\end{align*}
Thus, $\Mo_0$ need not be normalized to $\Mo_0(\R^2)=\id$, although we always have $\Mo_0(\R^2)\leq\id$.

The {\em weight at infinity} of the map $\Mo^{\rm av}$ is defined as 
\begin{align*}
\Mo^{\rm av} (\infty) & = \id -\sup\{ \Mo^{\rm av} (f)  \vert f\in C_c(\R^2),\, 0\leq f\leq 1\}\\
& = \id - \Mo_0(\R^2).
\end{align*}
Hence, the averaged covariant POVM $\Mo_0$ is normalized, and thus an observable, if and only if $\Mo^{\rm av}(\infty) = 0$. 
However, in our proof of Theorem \ref{thm:qp} the averaging does not lead to a normalized POVM, but instead the constructed map will have full weight at infinity, i.e., $\Mo^{\rm av}(\infty)=\id$. This just means that $\Mo_0(X)=0$ for all $X$, or that the ``measure part'' of  $\M^{\rm av}$ is zero. Let us now turn to the proof of Theorem \ref{thm:qp}.

\begin{proof}
Fix $0<\lambda\leq \jmd(\Qo,\Po)$ 
and let $\Mo$ be a joint observable for the corresponding noisy versions of $\Qo$ and $\Po$, i.e., 
\begin{align*}
\Mo(X\times\real) &= \lambda\Qo(X)  +(1-\lambda)\mu_1(X) \id  \, , \\
\Mo(\real\times Y) &=\lambda\Po(Y)  +(1-\lambda)\mu_2(Y) \id,
\end{align*}
where $\mu_1$ and $\mu_2$ are some probability measures. 
Let $\Mo^{\rm av}:C_b(\R^2)\to\lh$ be the averaged map constructed from $\Mo$ as explained earlier.
The margins of $\Mo^{\rm av}$ are defined in the obvious manner: For any $f\in C_b(\R)$ the functions $f_1(q,p) = f(q)$ and $f_2(q,p)=f(p)$ are in $C_b(\R^2)$, and we set $\Mo_i^{\rm av}(f)=\Mo^{\rm av}(f_i)$. Similarly, we can define the margins $m_i (f) = m(f_i)$ of the invariant mean $m$, which themselves turn out to be invariant means on $B(\R)$. Now, e.g., we have $\Mo[f_1] = \lambda\Qo[f]  +(1-\lambda)\mu_1[f] \id$, so that, for any state $\varrho$,
\begin{align*}
& \Theta[f_1;\varrho] (q,p) = \lambda\tr{\varrho W(q,p) \Qo[f^q] W(q,p)^* } \\
& \qquad \qquad + (1-\lambda) \mu_1[f^q] = \lambda \tr{\varrho\Qo[f]} + (1-\lambda) \mu_1[f^q]
\end{align*}
by Eq.~\eqref{eqn:pos_symmetry}. Denoting $(f*\mu_1^-)(q) = \mu_1[f^q]$, Eq.~\eqref{eq:defMav} then yields
$$
\Mo_1^{\rm av}(f) = \lambda \Qo[f] + (1-\lambda) m_1(f\ast\mu_1^-) \id .
$$ 

If $f\in C_c(\R)$,  then $f*\mu_1^-$ is a continuous function vanishing at infinity and hence $m(f*\mu_1^-)=0$ \cite[17.20]{AHAI63}. 
Hence, we have $\Mo_1^{\rm av}(f) = \lambda \Qo[f]$ for all $f\in C_c(\R)$, and by similar reasoning  $\Mo_2^{\rm av}(f) = \lambda \Po[f]$ for all $f\in C_c(\R)$. 
In other words, the unique POVMs corresponding to the margins of $\Mo^{\rm av}$ are scalar multiples of the position and momentum observables.

Now let us consider the margins of $\Mo_0$. For any $f\in C_c(\R)$ we have 
$\Mo_0[f_i] \leq  \Mo^{\rm av}(f_i)  = \Mo^{\rm av}_i(f) $ from which it follows that $\Mo_0(X\times \real)\leq \lambda \Qo(X)$ and $\Mo_0(\real \times Y)\leq \lambda \Po(Y)$.
In particular, $\Mo_0(X\times Y)\leq \lambda \Qo(X)$ and $\Mo_0(X\times Y)\leq \lambda \Po(Y)$.
The complementarity of $\Qo$ and $\Po$ then implies, in particular, that $\Mo_0(X\times Y)=0$ for all compact sets $X$ and $Y$.  Since $\real^2$ is $\sigma$-compact, we have $\Mo_0(\R^2)=0$ and thus $\Mo^{\rm av} (\infty)=\id$.

Consider next the weight at infinity of the margins. Since $m_i(f*\mu_i^-)=0$ for all $f\in C_c(\R)$, we have 
\begin{eqnarray*}
\Mo^{\rm av}_1 (\infty)
&=&\id-\sup\{ \lambda \Qo[f] \vert f\in C_c(\R), \, 0\leq f\leq 1 \}\\
&=&
(1-\lambda)\id, 
\end{eqnarray*}
and similarly
$\Mo^{\rm av}_2 (\infty) =  (1-\lambda)\id$.
However, we also have 
$$
\Mo^{\rm av}(\infty) \leq \Mo^{\rm av}_1(\infty) + \Mo^{\rm av}_2(\infty)
$$
(see the proof of \cite[Lemma 2]{Werner04}) so that 
$$
\id\leq (1-\lambda)\id + (1-\lambda)\id
$$
from which it follows that $\lambda\leq 1/2$. 
Since this is true for all $\lambda\leq\jmd(\Qo,\Po)$, we conclude that $\jmd(\Qo,\Po) = 1/2$. 
Therefore, $\Qo$ and $\Po$ are maximally incompatible.
\end{proof}

\section{Maximal incompatibility of number and phase}
As a second example we consider another pair of observables which is usually given the status of a complementary pair, namely, the quantum optical photon number and phase observables. Let $\hi$ be the Hilbert space spanned by the orthonormal basis $\{ \vert n \rangle \mid n\in\NAT_0 = \{0,1,2,\ldots  \}\}$ consisting of the number states and let $\No(\{n\})=\vert n\rangle\langle n\vert$ denote the number observable, i.e., the spectral measure of the number operator $N=\sum_{n=0}^\infty n\vert n\rangle\langle n \vert$. 
The canonical phase observable \cite{PSAQT82} is then defined as
\begin{equation*}
\Phase(X)  =\sum_{m,n=0}^\infty \frac{1}{2\pi} \int_X e^{i(m-n)\theta} \, {\rm d}\theta \, \vert m \rangle\langle n\vert
\end{equation*}
for all Borel sets $X\subseteq [0,2\pi)$. In particular,  $\Phase$ transforms covariantly under the phase shifts generated by the number operator, i.e., $e^{i\theta N} \Phase(X) e^{-i\theta N}= \Phase(X+\theta)$ where we regard $[0,2\pi)$ as a group with addition modulo $2\pi$. The number observable, on the other hand, is obviously phase shift invariant. 

The complementarity of $\No$ and $\Phase$ can be expressed by looking at their eigenstates and approximate eigenstates, respectively.
First, for a number state $\vert n\rangle$ the number distribution is peaked, but the phase distribution is uniform. 
Second, for coherent states $\vert z\rangle$ the canonical phase distribution approaches the delta distribution concentrated at $\arg(z)$ as $\vert z\vert\to \infty$ \cite{LaPe00}, while the number distributions get increasingly uniform, i.e., $\mo{\ip{z}{n}}^2\to 0$ as $\vert z\vert\to \infty$.
Using the complementarity and the specific structure of $\No$ and $\Phase$, we can prove the following result.

\begin{theorem}\label{thm:np}
The number and phase observables are maximally incompatible.
\end{theorem}

The core of the method for proving Theorem \ref{thm:np} is the same as for position and momentum, although some care needs to be paid to certain mathematical details. 
Again, before going into the details of the proof of this result, we will explain some general facts.

First of all, since the value space $\NAT_0$ of the number observable $\No$ is not a group but merely a semigroup, it is convenient to consider instead the extension $\No^{\rm ext}$ on $\integer$ obtained by setting $\No^{\rm ext}(\{ n\})=0$ for $n <0$. The important observation now is that $\jmd(\Phase, \No)\leq\jmd(\Phase,\No^{\rm ext})$. Indeed, if some noisy versions of $\Phase$ and $\No$ with a given $\lambda$ are jointly measurable, then by trivially extending the joint observable $\Mo$ into an observable $\Mo^{\rm ext}$ on $[0,2\pi)\times \integer$ we obtain a joint observable of noisy versions of $\Phase$ and $\No^{\rm ext}$ with the same  $\lambda$. 

Second, since $[0,2\pi)$ is a compact group, the averaging with respect to phase shifts can be done directly without using an invariant mean.
Indeed, suppose that $\Mo$ is a joint observable for noisy versions of number and phase, i.e., 
\begin{eqnarray}
\Mo(X\times \integer)  &=& \lambda \Phase(X) + (1-\lambda)\mu_1(X)\id,\label{eqn:phase_margin}\\
\Mo([0,2\pi)\times Y) &=& \lambda\No^{\rm ext}(Y)   + (1-\lambda)\mu_2(Y) \id.\label{eqn:number_margin}
\end{eqnarray}
Then, by defining 
$$
\Mo'(X\times Y) = \frac{1}{2\pi} \int e^{-i\theta N} \Mo((X+\theta)\times Y)e^{i\theta N}\, {\rm d}\theta
$$
we get an observable which satisfies $\Mo'((X+\theta)\times Y) = e^{i\theta N} \Mo'(X\times Y) e^{-i\theta N}$, i.e., it is phase shift covariant. The margins of $\Mo'$ differ from those of $\Mo$ only by the fact that the probability measure $\mu_1$ is replaced by the uniform distribution $u$ on $[0,2\pi)$. Therefore, without loss of generality we can always assume that the joint observable, if it exists, is phase shift covariant and hence the noise in the first margin is uniform.

One final difference when compared to the position-momentum case arises when we consider the generation of number shifts. Indeed, since $\Phase$ is not a spectral measure, we do not directly get a unitary representation as a suitable candidate for this. However, we can define for any $k\in\NAT_0$ the operator
\begin{equation*}
V(k)=\int e^{ik\theta} \, \Phase({\rm d}\theta) = \sum_{n=0}^\infty \vert n\rangle\langle n+k\vert,
\end{equation*}
so that the map $V:\NAT_0\to\lh$ is a (nonunitary) representation of the semigroup $\NAT_0$, which satisfies the commutation relation $e^{i\theta N}V(k) = e^{-ik\theta} V(k)e^{i\theta N}$. 
It is associated to number shifts as can be seen from the covariance condition $V(k) \No^{\rm ext}(Y) V(k)^* = \No^{\rm ext}(Y-k)$
Note  that this representation leaves the phase distribution invariant, i.e., $V(k)\Phase(X) V(k)^* = \Phase(X) $.  With this machinery, we are now ready to prove Theorem \ref{thm:np}.

\begin{proof}
Fix $0<\lambda \leq \jmd(\Phase, \No^{\rm ext})$ and let $\Mo$ be a phase shift covariant joint observable of noisy versions of $\Phase$  and $\No^{\rm ext}$ so that the margins of $\Mo$ are given by Eqs.~\eqref{eqn:phase_margin} and \eqref{eqn:number_margin} with the uniform noise in the first margin, i.e., $\mu_1=u$. 
 Now for any $f\in C_b([0,2\pi)\times \integer)$ and $k\in\NAT_0$  we set 
$$
\Theta[f;\varrho](k) = \tr{\varrho V(k) \Mo[ f^{(0,-k)}]V(k)^*}
$$
so that $\Theta[f;\varrho] \in B(\NAT_0)$. Hence, by defining 
$$
\tr{\varrho\Mo^{\rm av}(f)} = m(\Theta[f;\varrho]) 
$$
where $m$ is a semigroup invariant mean on $B(\NAT_0)$ with the property that $m(f) = \lim_{n\to +\infty} f(n)$ whenever this limit exists \cite[Thm.~17.5 and 17.20]{AHAI63}, we obtain a positive linear map $\Mo^{\rm av} : C_b([0,2\pi)\times \integer) \to\lh$  which satisfies the covariance condition $\Mo^{\rm av} (f^{(\theta,k)}) = e^{-i\theta N} V(k) \Mo^{\rm av} (f) V(k)^* e^{i\theta N} $.

Using the number shift invariance of the noisy phase observable, we see that $\Mo^{\rm av}_1(f)=\lambda \Phase[f] + (1-\lambda) u[f]\id$ so that $\Mo^{\rm av}_1(\infty)=0$, and the same argument as in the proof of Theorem \ref{thm:qp} shows that $\Mo^{\rm av}_2(f)=\lambda \No^{\rm ext}[f]$ for all $f\in C_c(\integer)$. In particular, if $\Mo_0$ again denotes the POVM on $[0,2\pi)\times \integer$ corresponding to the restriction of $\Mo^{\rm av}$ to $C_c([0,2\pi)\times \integer)$, then 
\begin{align}
\Mo_0 (X\times\{n\}) & \leq \Mo_0 ([0,2\pi)\times\{n\}) = \Mo^{\rm av} (\chi_{[0,2\pi)\times\{n\}}) \nonumber\\
& = \Mo^{\rm av}_2 (\chi_{\{n\}}) = \lambda \No^{\rm ext}(\{n\}) \,,\label{eq:MavPhi}
\end{align}
where $\chi_E$ denotes the indicator function of a set $E$. Note that in the first and third equality we have used the facts that $\chi_{[0,2\pi)\times\{n\}}\in C_c ([0,2\pi)\times \integer)$ and $\chi_{\{n\}}\in C_c ( \integer)$, respectively. In particular, $\Mo_0 (X\times\{n\})=0$ for all $n<0$ and for $n\geq 0$ we have $\Mo_0 (X\times\{n\})\leq \lambda \vert n\rangle\langle n\vert$. It follows that there exists a number $0\leq \omega(X,n)\leq 1$ such that 
\begin{equation}\label{eqn:np_radon}
\Mo_0(X\times\{n\})=\omega(X,n)\lambda\No^{\rm ext}(\{n\}).
\end{equation}
Note that $\omega(X,n)=0$ for all $n<0$ but for $n\geq 0$ the map $\omega(\cdot,n)$ is actually a positive measure. By the covariance of $\Mo^{\rm av}$ we have
$$
e^{-i\theta N} V(k) \Mo_0(X\times Y) V(k)^*e^{i\theta N} = \Mo_0((X-\theta)\times (Y-k)) 
$$
so that by applying this to Eq.~\eqref{eqn:np_radon} we get
$$
\omega(X,n)\lambda\vert n-k\rangle\langle n-k\vert = \omega(X-\theta,n-k)\lambda\vert n-k\rangle\langle n-k\vert
$$
for all $n\geq k\geq 0$.
Hence, we have $\omega(X-\theta,n-k)=\omega(X,n)$ so by uniqueness of the Haar measure on $[0,2\pi)$ there exists a positive constant $c$, independent of $n$, such that $\omega(X,n)=c\,u(X)$ for all $n\geq 0$. Eq.~\eqref{eqn:np_radon} then gives us $ \Mo_0 ([0,2\pi)\times \integer) = c \lambda \id$. On the other hand, we know by Eq.~\eqref{eq:MavPhi} that  $\Mo_0 ([0,2\pi)\times \{n \})= \lambda \No^{\rm ext}(\{ n \})$ so that $\Mo_0 ([0,2\pi)\times \integer) =\lambda \id$. By comparison, we have $c=1$ and thus 
$$
\Mo_0(X\times\{n\})=\lambda u(X) \No^{\rm ext} (\{ n\}).
$$


Therefore, for any $f\in C_b([0,2\pi))$ we have $\Mo_0[f_1] = \lambda u[f]$ and the inequality $\Mo_0[f_1] \leq \Mo^{\rm av}_1(f) = \lambda \Phase[f]  + (1-\lambda) u[f] \id$  implies that
\begin{equation*}
\lambda \Phase(X)  + (1-2\lambda) u(X) \id\geq 0
\end{equation*}
for all Borel sets $X\subseteq [0,2\pi)$.
Since for coherent states $\vert z\rangle$ the canonical phase distribution approaches the delta distribution concentrated at $\arg(z)$ as $\vert z\vert\to \infty$, by setting $\arg(z)=0$ and $X=[\pi/2,\pi]$ we obtain
\begin{equation*}
0\leq \lim_{\vert z\vert\to\infty} \left( \lambda \langle z\vert \Phase(X)\vert z\rangle  + (1-2\lambda) u(X) \right) = (1-2\lambda )\frac{1}{4} \, .
\end{equation*}
This means that $\lambda\leq \frac{1}{2}$, so that $\jmd(\Phase, \No^{\rm ext})=\frac{1}{2}$ and thus also $\jmd(\Phase, \No)=\frac{1}{2}$.
\end{proof}

\section{Summary and outlook}

The concepts of joint measurability region and joint measurability degree are ways of quantifying the incompatibility of two observables in any probabilistic theory. One can even go a step further and take the joint measurability region or degree of the most incompatible pair of observables in a given theory to describe the degree of incompatibility inherent in the theory \cite{BuHeScSt13}. Therefore, in order to gain a better understanding of the incompatibility inherent in quantum theory as compared to other probabilistic theories, we need to have better knowledge of maximally incompatible quantum observables.

Using a quantum cloning device as a means of performing approximate joint measurements, we have derived a dimension dependent lower bound for the smallest possible joint measurability degree in a finite dimensional Hilbert space. For any finite dimension this bound is strictly greater than the joint measurability degree of a maximally incompatible pair, and therefore our result shows that in quantum theory maximal incompatibility requires an infinite dimensional Hilbert space. What still remains an open question is whether or not in a fixed finite dimension the corresponding bound can actually be reached by some pair of observables. Indeed, one might expect that two canonically conjugated observables would be as incompatible as any two observables can be, but as we have demonstrated, their joint measurability degree never coincides with the derived lower bound. Therefore we only know that the joint measurability degree of the most incompatible pair lies somewhere between these two values. 

In the case of an infinite dimensional Hilbert space we have shown that two of the most common pairs of complementary observables (position and momentum; number and phase) constitute maximally incompatible pairs. In both cases the complementarity is explicitly used in the proof, and therefore it is natural to ask if there is in general some connection between maximal incompatibility and other formulations of the incompatibility of observables. We leave this as a possible topic for future investigations.

\section{Acknowledgement}
T.H. acknowledges financial support from the Academy of Finland (grant no. 138135).  J.S. and A.T. acknowledge financial support from the Italian Ministry of Education, University and Research (FIRB project RBFR10COAQ). M.Z. acknowledges support of VEGA 2/0127/11, GACR P202/12/1142 and COST Action MP1006.


\bibliographystyle{plain}
\bibliography{bibliography}

\end{document}